\documentclass[twocolumn]{aastex631}

\usepackage{bm}
\usepackage{amsmath}
\usepackage{hyperref}

\begin{document}

\title{Low-energy Injection and Nonthermal Particle Acceleration in Relativistic Magnetic Turbulence}

\author[0000-0001-7235-392X]{Divjyot Singh}
\thanks{dsingh@u.northwestern.edu}
\affiliation{Los Alamos National Laboratory, Los Alamos, NM 87545, USA}
\affiliation{Department of Engineering Sciences \& Applied Mathematics, Northwestern University,
2145 Sheridan Road, Evanston, IL, 60208, USA}
\affiliation{Center for Interdisciplinary Exploration and Research in Astrophysics (CIERA), Northwestern University, 1800 Sherman Ave, Evanston, IL 60201, USA}

\author[0000-0001-6155-2827]{Omar French}
\thanks{National Science Foundation Graduate Research Fellow}
\affiliation{Los Alamos National Laboratory, Los Alamos, NM 87545, USA}
\affiliation{Center for Integrated Plasma Studies, Department of Physics, 390 UCB, University of Colorado, Boulder, CO 80309, USA}

\author[0000-0003-4315-3755]{Fan Guo}
\affiliation{Los Alamos National Laboratory, Los Alamos, NM 87545, USA}
\affiliation{New Mexico Consortium, Los Alamos, NM 87544, USA}

\author[0000-0001-5278-8029]{Xiaocan Li}
\affiliation{Los Alamos National Laboratory, Los Alamos, NM 87545, USA}
 
\begin{abstract}

Relativistic magnetic turbulence has been proposed as a process for producing nonthermal particles in high-energy astrophysics. The particle energization may be contributed by both magnetic reconnection and turbulent fluctuations, but their interplay is poorly understood. It has been suggested that during magnetic reconnection the parallel electric field dominates the particle acceleration up to the lower bound of the power-law particle spectrum, but recent studies show that electric fields perpendicular to the magnetic field can play an important, if not dominant role. In this study, we carry out two-dimensional fully kinetic particle-in-cell simulations of magnetically dominated decaying turbulence in a relativistic pair plasma. For a fixed magnetization parameter $\sigma_0 = 20$, we find that the injection energy~$\varepsilon_{\rm inj}$ converges with increasing domain size to~$\varepsilon_{\rm inj} \simeq 10 \, m_ec^2$. In contrast, the power-law index, the cut-off energy, and the power-law extent increase steadily with domain size. We trace a large number of particles and evaluate the contributions of the work done by the parallel ($W_\parallel$) and perpendicular ($W_\perp$) electric fields during both the injection phase and the post-injection phase. We find that during the injection phase, the $W_\perp$ contribution increases with domain size, suggesting that it may eventually dominate injection for a sufficiently large domain. In contrast, on average, both components contribute equally during the post-injection phase, insensitive to the domain size. For high energy ($\varepsilon \gg \varepsilon_{\rm inj}$) particles, $W_\perp$ dominates the subsequent energization. These findings may improve our understanding of nonthermal particles and their emissions in astrophysical plasmas.

\end{abstract}

\section{Introduction}\label{sec:intro}

Magnetic turbulence in plasmas reveals itself through fluctuating magnetic fields, bulk velocity, and density over a broad range of spatial and temporal scales. It is commonly found and studied in astrophysical environments such as pulsar wind nebulae \citep{Porth2014, Lyutikov2019,Cerutti_2020,Lu2021}, stellar coronae and flares \citep{Matthaeus1999,Cranmer2007,liu2006,Fu2020,Peera2021}, black hole accretion disks \citep{Balbus1998,Brandenburg2005,Sun2021}, radio lobes \citep{Vogt2005,O_Sullivan2009}, and jets from active galactic nuclei \citep{Marscher2008,Zhang2023}. All of these systems exhibit high-energy emissions that suggest nonthermal particle acceleration. In turbulent plasmas, the kinetic energy from large-scale motion cascades to smaller and smaller scales, which is eventually dissipated through turbulence-particle interactions. Understanding how particles in turbulent plasmas get accelerated to high energy is an unsolved problem in high-energy astrophysics.

Turbulence is often invoked as a particle acceleration mechanism that leads to nonthermal particle spectra. Recently, several studies have used kinetic particle-in-cell (PIC) simulations to gain insight into nonthermal particle acceleration mechanisms in its relativistic regime \citep{Zhdankin_2017, Zhdankin_2018, Comisso2018, Comisso2019, Wong_2020, Hankla_2021, Vega_2022}. The most commonly discussed acceleration mechanism in magnetic turbulence is stochastic Fermi acceleration \citep{Fermi_1949,Petrosian2012,Lemoine2020}, where particles can gain energy by scattering back and forth in the turbulent fluctuations. Magnetic reconnection \citep{Biskamp2000,Zweibel2009,Yamada2010,Ji2022,Yamada2022}, which occurs naturally as magnetic turbulence generates thin current sheets, may also support strong particle acceleration \citep{Sironi2014,Guo2014,Guo2015,Werner2016,Guo2020}. More interestingly, magnetic reconnection can have an intriguing relation with turbulence and their interplay during particle acceleration is not completely clear \citep{Loureiro2017,Dong2018,Dong2022,Comisso2019,Li2019,Zhang2021,Zhang2024,Guo2021}. Nevertheless, these recent numerical simulations and theoretical models suggest that magnetic turbulence, especially in its relativistic limit ($\sigma \equiv B^2/4\pi h \gg 1$; i.e. the magnetic enthalpy $B^2/4\pi$ greatly exceeds the plasma enthalpy $h$), plays a major role in nonthermal particle acceleration.

In general, Fermi acceleration requires particle injection mechanism(s) to accelerate particles to energies that enable them to participate in a continual acceleration process. This process naturally defines an \textit{injection energy}, beyond which injected particles enter the power-law range of the particle spectrum \citep{French_2023}. The injection problem has recently been studied in the context of relativistic magnetic reconnection \citep{Guo2019,Ball2019,Kilian2020,Sironi2022,French_2023,Guo2023}. While it has been suggested that during magnetic reconnection the parallel electric field~$\textbf{E}_\parallel \equiv (\textbf{E} \cdot \textbf{B}) \textbf{B}/ |\textbf{B}|^2$ dominates the injection \citep{Ball2019}, studies have shown that perpendicular electric fields ($\textbf{E}_\perp \equiv \textbf{E} - \textbf{E}_\parallel$) can play an important, if not dominant role \citep{Kilian2020,French_2023}. Meanwhile, X-points with $|E|>|B|$ are shown to be negligible for particle injection and high-energy acceleration \citep{Guo2019,Guo2023}. Particle injection has also been investigated in relativistic magnetic turbulence \citep{Comisso2019}, where parallel electric fields in reconnection diffusion regions were concluded to dominate the injection process. Meanwhile, the subsequent particle energization in the power law was shown to be dominated by perpendicular electric fields ($\textbf{E}_\perp$) from stochastic scattering off turbulent fluctuations. However, \citet{Comisso2019} focused only on a small population of high energy particles with final energies many times greater than the injection energy. Since the importance of~$\textbf{E}_\perp$ has been demonstrated in magnetic reconnection, it is worthwhile to investigate whether~$\textbf{E}_\perp$ is important in magnetic turbulence as well. 

In a recent study, \citet{French_2023} analyzed particle injection and further acceleration in relativistic magnetic reconnection with emphasis on the influence of guide field and domain size. They measured the injection energy of each nonthermal particle spectrum using a spectral fitting procedure. They decompose the work done by parallel and perpendicular electric field components and quantify the contributions by different mechanisms, thereby illuminating which mechanism dominates the initial energization and the subsequent nonthermal acceleration. In this paper, we employ a similar methodology to study collisionless relativistic turbulence by carrying out two-dimensional (2D) PIC simulations and calculating the shares of work done by parallel ($W_\parallel$) and perpendicular ($W_\perp$) electric fields. We find that, similar to magnetic reconnection, the contribution of~$W_\perp$ to particle injection grows with increasing domain size until the largest simulation domain, and may all exceed $50\%$ contribution for macroscale systems. However, in contrast to magnetic reconnection, the relative contributions of~$W_\parallel$ vs~$W_\perp$ to subsequent energization of particles of energies~$\varepsilon > \varepsilon_{\rm inj}$ is relatively insensitive to domain size. 

The rest of the paper is organized as follows: Section \ref{sec:simulation_setup} describes our simulation setup. In Section \ref{sec:results} we present the simulation results and analyses for understanding the particle injection and nonthermal particle acceleration. Section \ref{sec:discussions_conclusion} discusses implications for observations and summarize the conclusions.

\section{Numerical Simulations} \label{sec:simulation_setup}

We use the Vectorized Particle-In-Cell ({\sc VPIC}) simulation code to investigate nonthermal particle acceleration in relativistic magnetic turbulence. VPIC solves the relativistic Maxwell-Vlasov equations to self-consistently evolve kinetic plasmas and their interaction with electromagnetic fields \citep{Bowers2008a, Bowers2008b, Bowers2009}. 
We simulate magnetically-dominated decaying turbulence in a two-dimensional (2D) square domain ($x$-$y$) of size $L^2$. The initial setup is similar to earlier work \citep{Comisso2019,Peera2021,Zhang2023}, where an electron-positron pair plasma is initialized with a turbulent magnetic field $\bm{B} = B_0 \bm{\hat{z}} + \delta \bm{B}$. $B_0$ is the magnitude of the uniform component and $\delta \bm{B}$ is the fluctuating component, which is given by 
\begin{equation}
\label{eq:B_perturbation}
    \delta\bm{B} (\bm{x}) = \sum_{\bm{k}}  \delta B ( \bm{k} ) \bm{\hat{\xi}}( \bm{k})\exp [i \left( \bm{k} \cdot \bm{x} + \phi_{\bm{k}} \right)]
\end{equation}
Here, $\delta B ( \bm{k} )$ is the Fourier amplitude of the mode with wavevector $\bm{k}$, $\bm{\hat{\xi}}( \bm{k} ) = i \bm{k} \times \bm{B}_0 / |\bm{k} \times \bm{B}_0|$ are the Alfv\'enic polarization unit vectors, and $\phi_{\bm{k}}$ expresses random phases. $\bm{k}$ represents the wavevector such that $\bm{k} = (k_x, k_y)$, where $k_x = 2 m \pi/ L$ and $k_y = 2 n \pi / L$ with $m \in \{ -N, \dots, -1, 1, \dots, N\}$ and $n \in \{ -N, \dots, -1, 1, \dots, N\}$.  $N$ is the number of modes along each dimension, which is set to be~$8$ in this paper. We also define wavenumber $k = |\bm{k}| = \sqrt{k_x^2 + k_y^2}$ as the amplitude of the wavevector. The boundary conditions are periodic for both particles and fields. The initial electric field $\textbf{E}$ is set to 0.

\begin{figure*}[ht!]
\includegraphics[width=\textwidth]{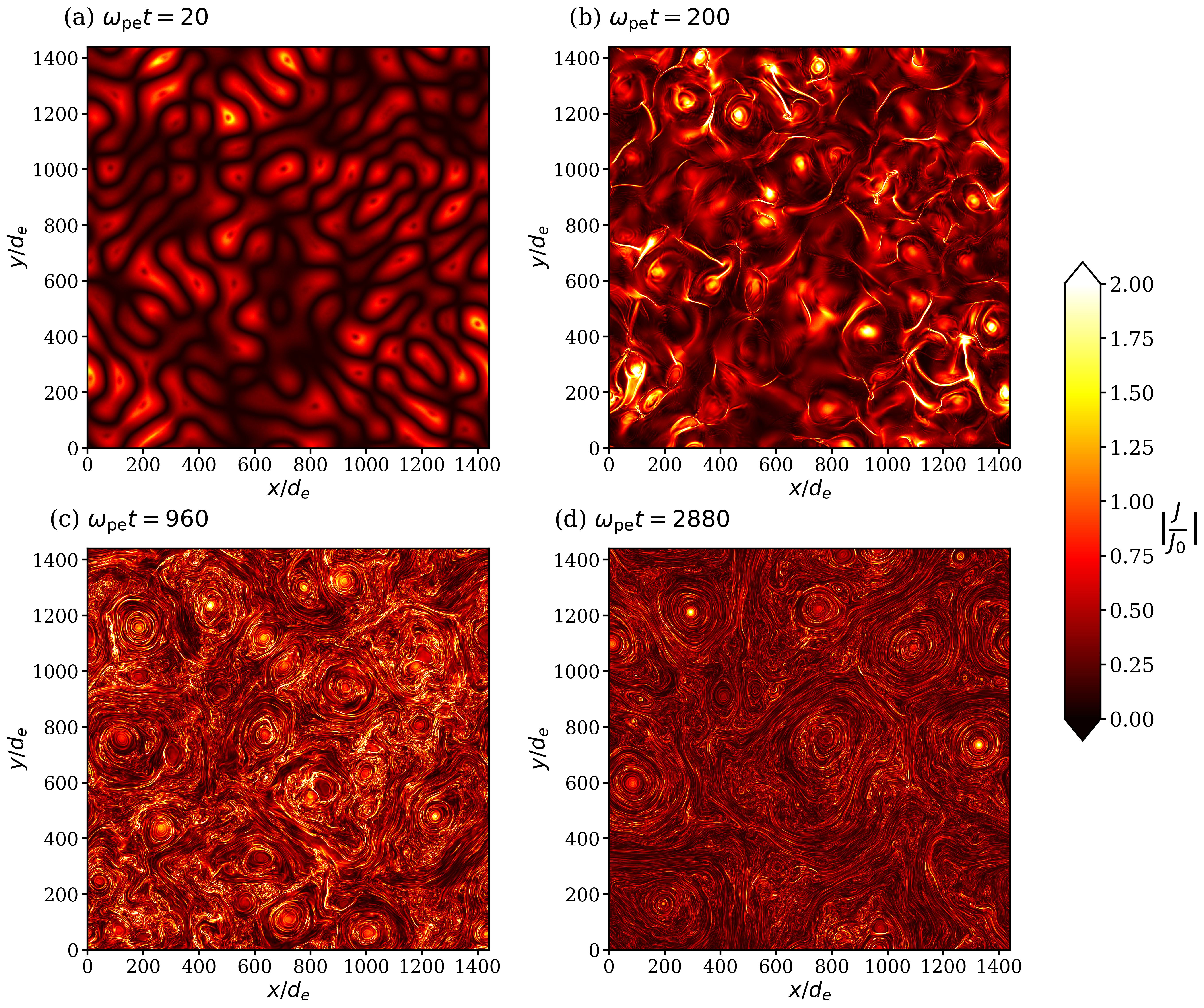}
\caption{Current density magnitude $|J/J_0|$ of the case $L/d_e = 1440$ at times $ \omega_{\rm pe} t = $ (a) $20$, (b) $200$, (c) $960$, and (d) $2880$. An animation is also available  on YouTube \url{https://youtu.be/NB4ulJ39H5M} which shows the evolution of current density from $ \omega_{\rm pe} t = 20$ to $2880$ in steps of $20$.}
\label{fig:abs_j}
\end{figure*}

We initialize the plasma and magnetic fields with magnetization parameter $\sigma_0 \equiv B_0^2/(4 \pi n_0 m_e c^2) = \omega_{\rm ce}^2/2\omega_{\rm pe}^2 = 20$, where~$\omega_{\rm pe} \equiv \sqrt{4\pi n_e e^2/m_e}$ is the plasma electron frequency and $\omega_{\rm ce} \equiv eB_0/m_e c$ is the electron cyclotron frequency defined using the uniform background magnetic field $B_0$. Here, $m_e$ is the electron mass, $c$ is the speed of light, $e$ is the electron charge, and $n_0 = n_p + n_e$ is the number density of the pair plasma in the simulation domain. The turbulence amplitude ${\delta B}_{\rm rms 0}/B_0=1$, where ${\delta B}_{\rm rms 0}$ is the space-averaged root-mean-square value of the initial magnetic field fluctuations. The domain size~$L$ is normalized by the electron skin depth~$d_e \equiv c/\omega_{\rm pe}$ and each~$d_e$ is resolved to 4 grid cells (i.e., $d_e = 4 \Delta x$). To allow most of the turbulent magnetic energy to be converted to the particles, the simulations are run for two light crossing times~$2L/c$. To independently examine the influence of domain size on our results, we run an array of otherwise identical simulations with $L/d_e \in \{512, 1024, 1440, 2048, 2880, 4096\}$.

In all our simulations, we use 100 particles of each species per cell that are initialized with a Maxwellian distribution with dimensionless temperature~$\theta_0 \equiv k_B T_0/m_e c^2 = 0.3$. Here, $k_B$ is the Boltzmann constant and $T_0$ is the initial plasma temperature. We also have done some test simulations with a larger number of particles per cell and/or higher spatial resolution and found that the results described below still hold.

For each simulation, we trace $\sim 200,000$ particles of each species and save the electric and magnetic fields $\bm{E}$ and $\bm{B}$ as well as velocities~$\textbf{v}$ at their positions at every time step, to understand their injection and nonthermal particle acceleration \citep{Li2023}. 

\section{Simulation Results} \label{sec:results}

\begin{figure*}[ht!]
\includegraphics[width=0.8\textwidth]{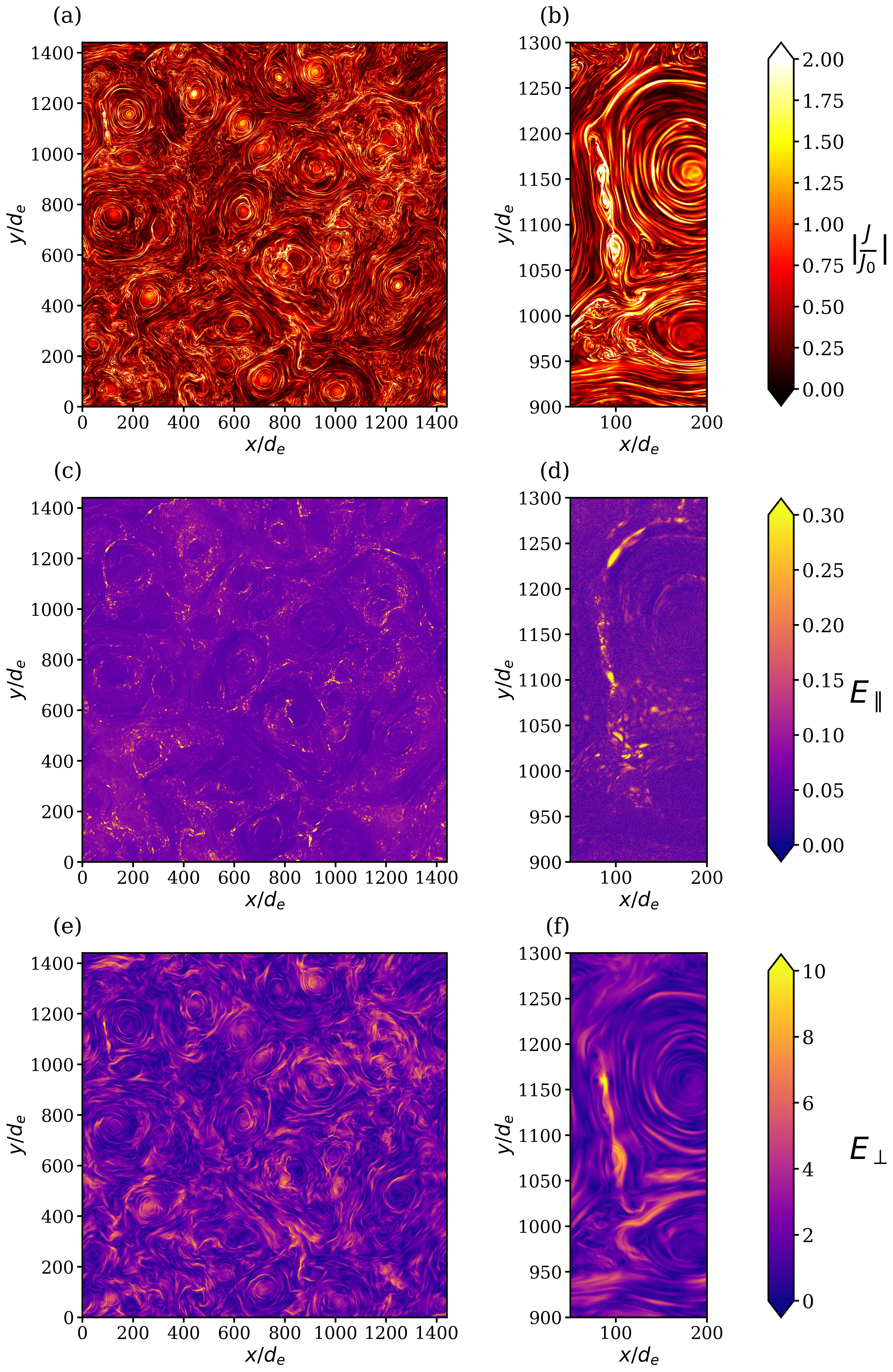}
\caption{Color maps of (a, b) current density magnitude ($|J/J_0|$), (c, d) parallel electric field ($E_\parallel$), and (e, f) perpendicular electric fields ($E_\perp$) for $L/d_e = 1440$ when~$\omega_{\rm pe} t = 960$. The right column [panels~(b, d, f)] are zoomed-in versions of the left column [panels~(a, c, e)] that focus on a specific reconnection region around~$x/d_e = 100$, $y/d_e = 1100$.}
\label{fig:reconnection_J_fields}
\end{figure*}

\begin{figure}[ht!]\includegraphics[width=0.5\textwidth]{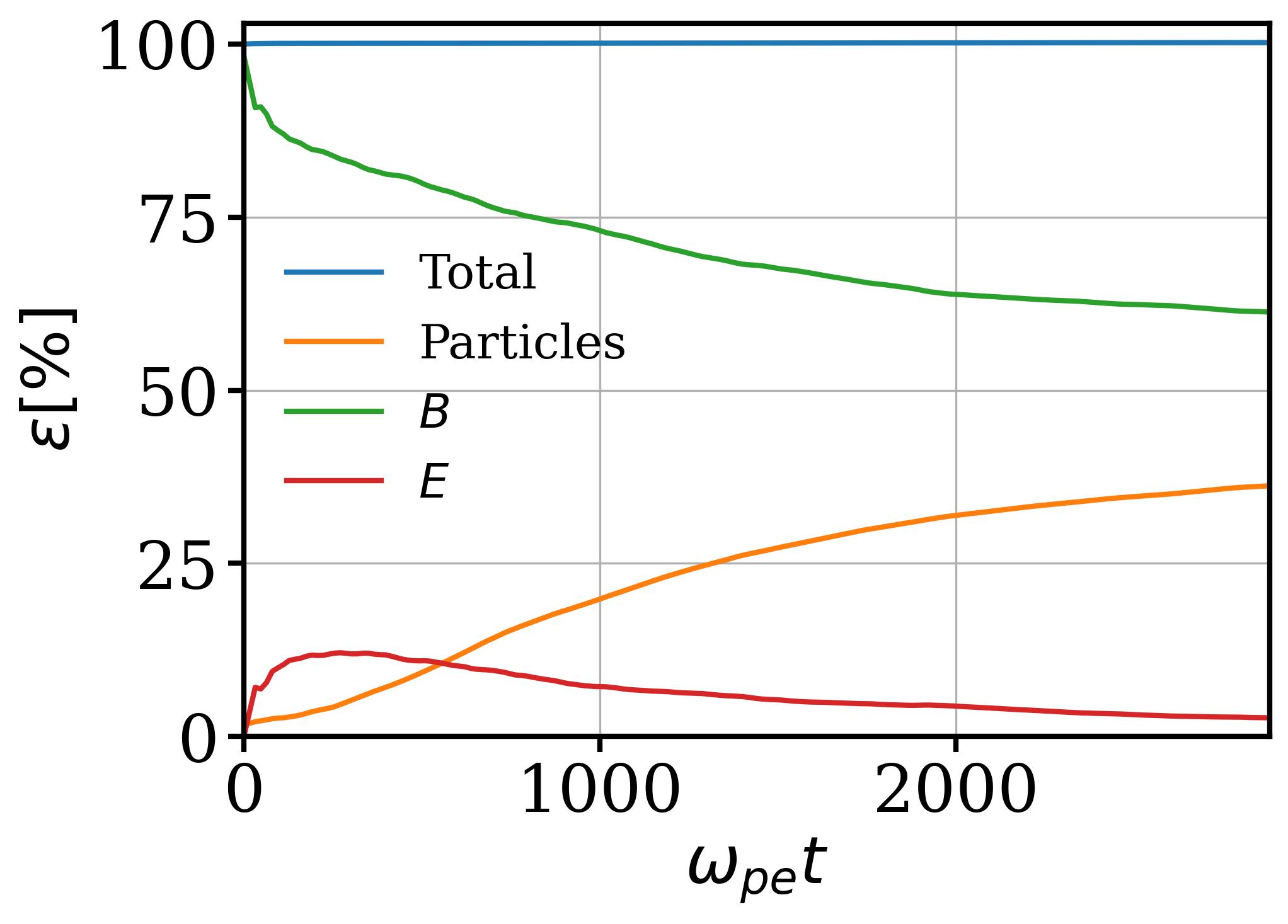}
\caption{Evolution of the percentage of total energy stored in the particles, magnetic fields, and electric fields in the standard run with $L/d_e = 1440$.} 
\label{fig:total_energy}
\end{figure}

\begin{figure}[ht!]\includegraphics[width=0.5\textwidth]{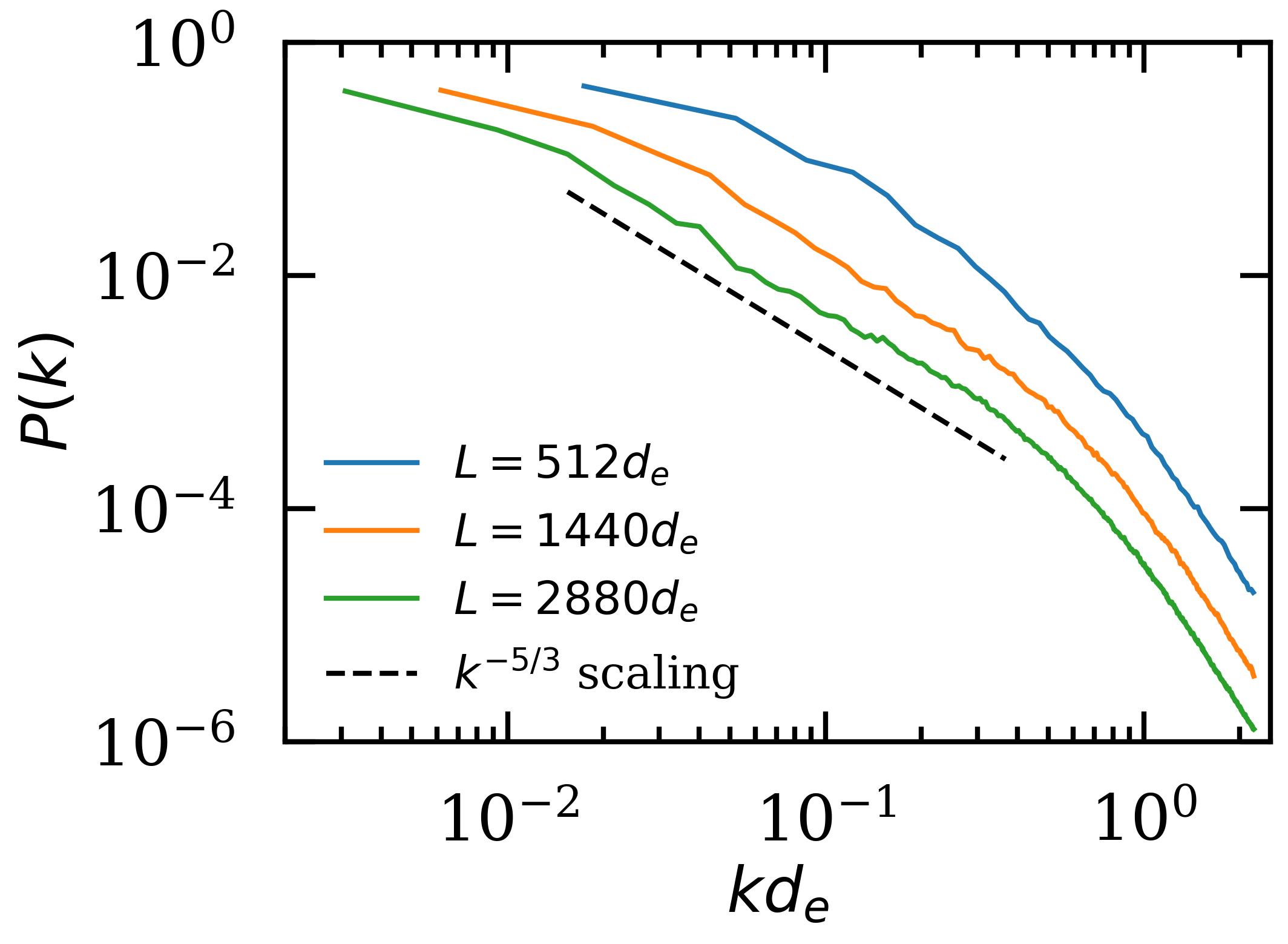}
\caption{
Power spectra of magnetic field fluctuations normalized with the total fluctuating power as a function of wavenumber $k$ for different domain sizes at $t \simeq 2 L/c$.} 
\label{fig:power_spectra}
\end{figure}

\begin{figure}[ht!]\includegraphics[width=0.5\textwidth]{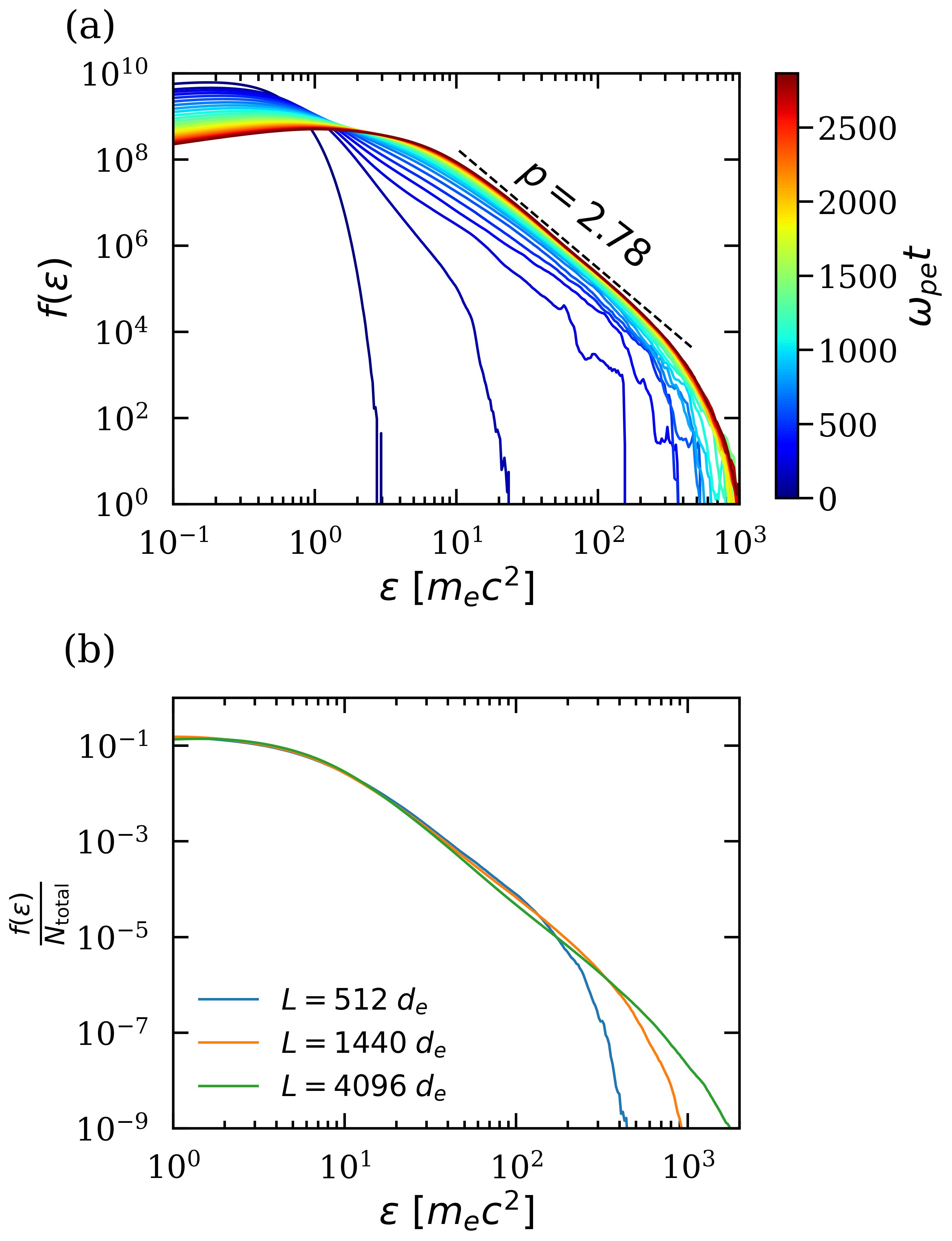}
\caption{(a) Time evolution of the particle energy spectrum for~$L/d_e = 1440$. The dashed line represents the slope of the fully evolved spectrum. (b) Normalized particle energy spectra at final times for different domain sizes.}
\label{fig:spectra}
\end{figure}

Figure~\ref{fig:abs_j} shows the evolution of the magnitude of electric current density $|J/J_0|$ in the simulation domain for the simulation with $L/d_e = 1440$ at times $\omega_{\rm pe} t = $ (a) $20$, (b) $200$, (c) $960$, and (d) $2880$, normalized to~$J_0 \equiv n_0 e c/2$.   
The initial perturbation seen in panel~(a) generates fluctuations across different scales, after a brief initial phase. As turbulence develops, many plasmoids\footnote{Note that many of the large plasmoids are due to the initial evolution of the initial perturbation, whereas during the evolution of the simulation small-scale plasmoids are generated during the reconnection process, which indicates that the energy is transferred to smaller scales.} and current sheets are produced in 2D turbulence, where magnetic reconnection is likely to happen (panel b).

Figure~\ref{fig:reconnection_J_fields} zooms in on a reconnection site occurring in the simulation at~$\omega_{\rm pe}t = 960$ and displays colormaps of (a-b) the absolute current density~$\lvert J/J_0 \rvert$, (c-d) the parallel electric field~$E_\parallel$, and (e-f) the perpendicular electric field~$E_\perp$. Here, $E_\parallel$ and $E_\perp$ are plotted in units of $B_0/\sqrt{2\sigma_0}$. From inspecting these figures we see that~$E_\perp \gg E_\parallel$ on a global scale, and it becomes clear that~$E_\parallel$ is well-localized to reconnection X-points at plasmoid interfaces. However, $E_\perp$ can still have a substantial strength at reconnection regions owing to the reconnection outflow immediately downstream of these X-points \citep{French_2023}.

In Figure~\ref{fig:total_energy}, we show how the fractions of energy stored in particles, magnetic fields, and electric fields evolve as the simulation proceeds. The total energy is well conserved. As the turbulence decays and reconnection events begin liberating magnetic field energy into nearby particles, the fraction of energy stored by particles grows from~$\sim 2.5\%$ at~$t = 0$ to~$\sim 35\%$ by the final time. This corresponds to the decrease of magnetic field energy. Since the initial electric field is set to be zero and induced rapidly due to the changing magnetic field, its energy experience a strong, transit growth in the initial stage $\omega_{\rm pe} t<500$.

Figure~\ref{fig:power_spectra} shows the power spectra of magnetic field fluctuations $\delta \bm{B}$ for various domain sizes at 2 light crossing time. The power $P(k)$ is normalized by the total power for that simulation at that time. In all the cases, we observe that a Kolmogorov-like $k^{-5/3}$ scaling quickly established and last until the end of the simulation.  For larger domains, the fluctuations extends to larger spatial scales (lower $k$), and the small scale fluctuations have lower amplitude. Meanwhile, the amplitude of the fluctuation $\delta B_{rms}$ decays from 1.0 to about 0.5 in the end of the simulation, quite consistently in all simulations.

\begin{figure*}[ht!]\includegraphics[width=\textwidth]{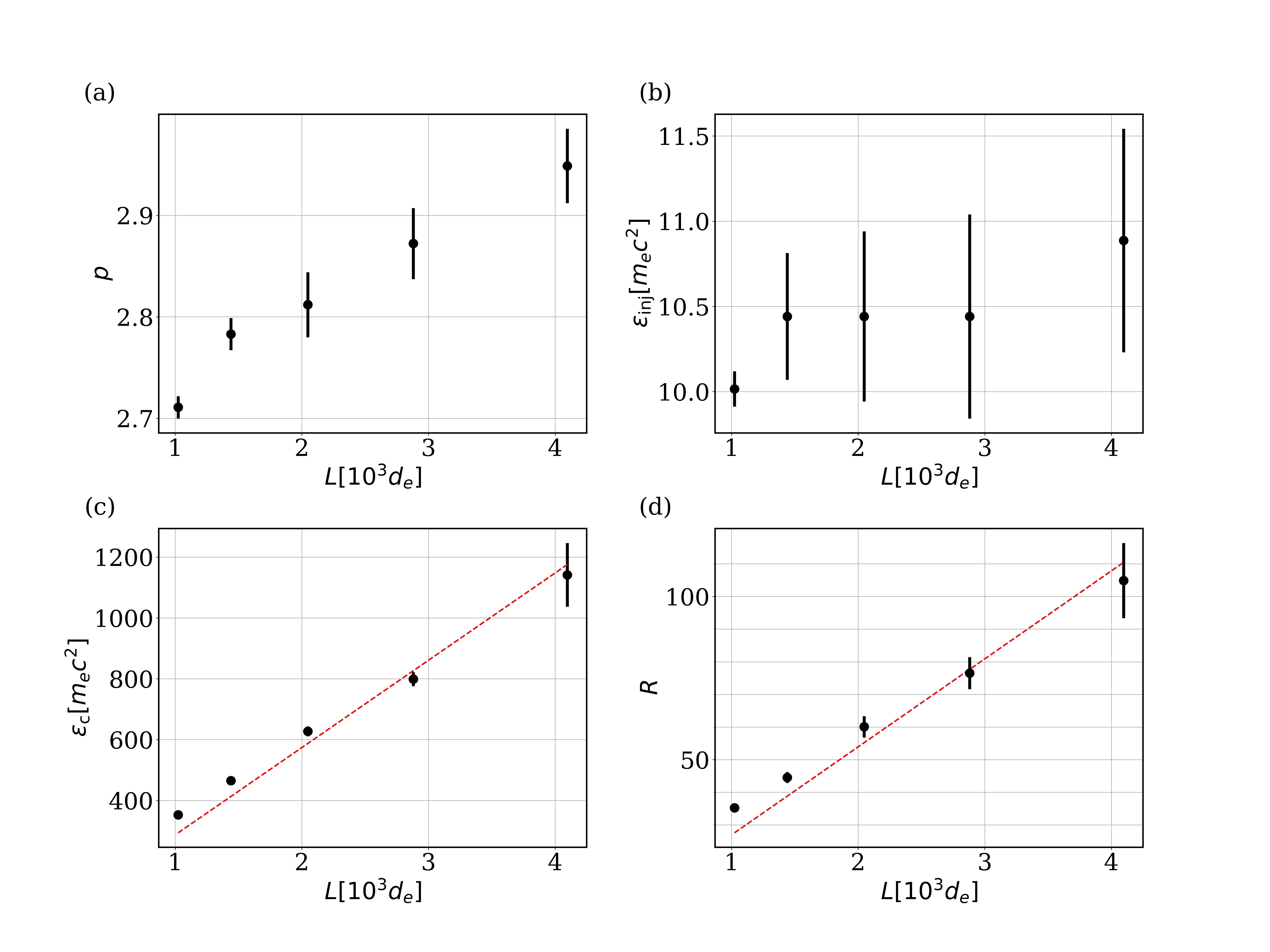}
\caption{(a) Power law index $p$, (b) Injection energy $\varepsilon_{\rm inj} [m_e c^2]$, (c) cutoff energy $\varepsilon_c [m_e c^2]$, and (d) power-law extent $R \equiv \varepsilon_c/\varepsilon_{\rm inj}$ for different domain sizes. The red dashed lines show the linear fits (c) $\varepsilon_c/(m_ec^2)=286.92(L/10^3d_e)$ and (d) $R = 26.96(L/10^3d_e)$.}
\label{fig:spectrum_params}
\end{figure*}

We analyze the nonthermal spectra for all of our simulations, and quantify several spectral features: power-law index~$p$ that represents the slope in the nonthermal region of the spectrum, the injection energy~$\varepsilon_{\rm inj}$ and cut-off energy~$\varepsilon_c$ that mark the lower and upper energy bounds of the nonthermal region respectively, and the power-law extent~$R\equiv \varepsilon_c/\varepsilon_{\rm inj}$. From these nonthermal particle spectra, we perform a fitting procedure at the end of the simulation to obtain the characteristic parameters ($\varepsilon_{\rm inj}$, $\varepsilon_c$, $p$) of our particle spectra \citep{Werner_2017, French_2023}, from which we also calculate the power-law extent~$R$. The procedure begins by smoothing a particle spectrum~$f$ via isotonic regression so that the local power-law index $p_\varepsilon \equiv -\,d \log{f(\varepsilon)}/d \log{\varepsilon}$ can be defined. Here, $\varepsilon$ refers to the particle energy. Then all ``valid" power-law segments are obtained by brute force, where validity is determined by a predefined power-law tolerance and minimum power-law extent, yielding a list of power-law indices, injection energies, and cutoff energies (see \cite{French_2023} for details). Finally, after removing duplicates (e.g., identical power-law segments resulting from different~$p$-tolerances) and outliers (i.e., data points beyond~$\pm$ 2 standard deviations from the mean) from each collection of values, each characteristic parameter ($p$, $\varepsilon_{\rm inj}$, $\varepsilon_c$) is defined by the mean of its collection and its error by one standard deviation of its collection. 

\begin{figure*}[ht!]\includegraphics[width=\textwidth]{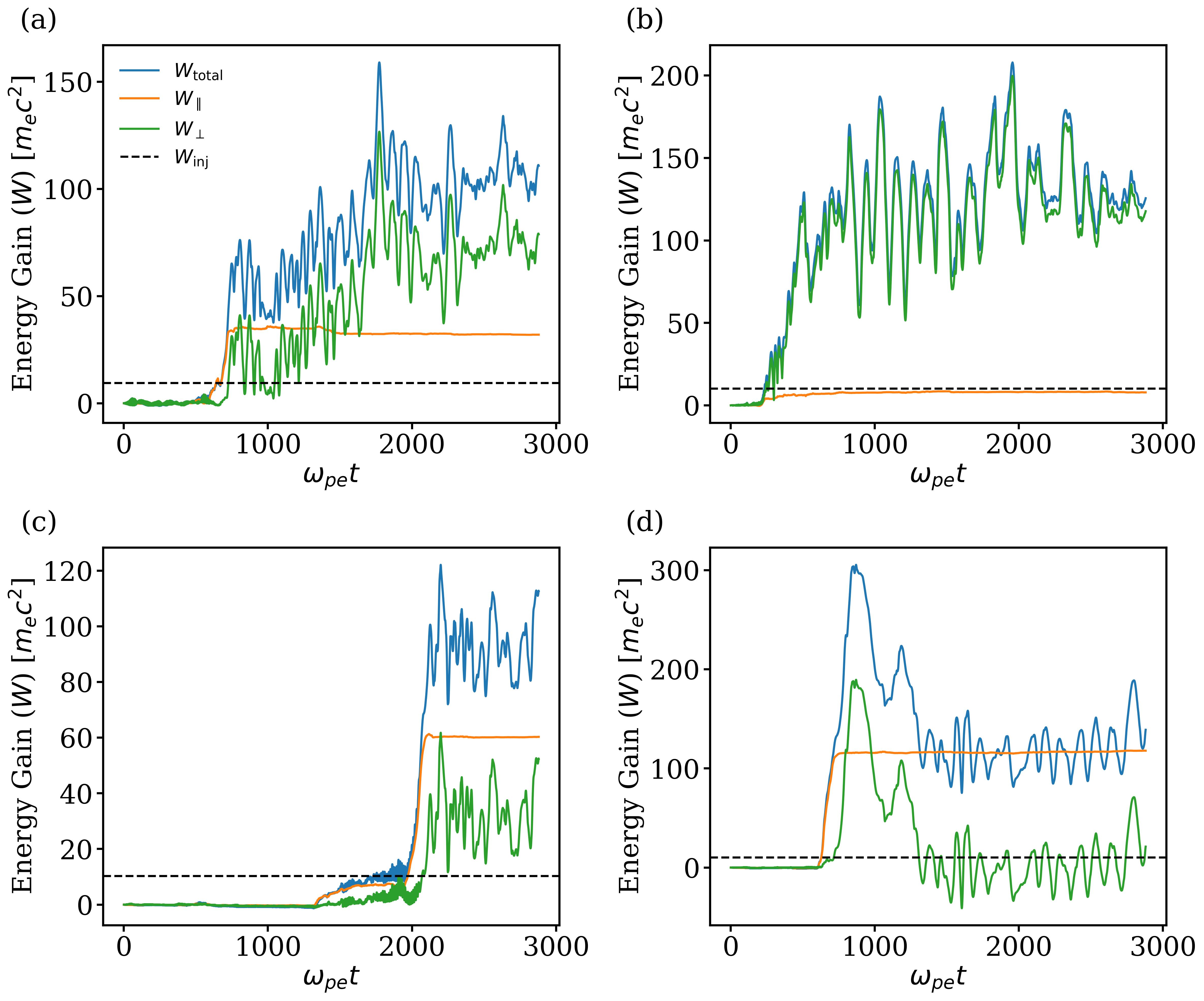}
\caption{Contributions to total energy gain by $W_\parallel$ and $W_\perp$ for four tracer particles with final energies (a) $112 \, m_e c^2$, (b) $127 \, m_e c^2$, (c) $115 \, m_e c^2$, and (d) $139 \, m_e c^2$. The black dashed line represents the injection threshold $W_{\rm inj} \equiv \varepsilon_{\rm inj} - \varepsilon_0$ and has the values (a) $9.5 \, m_e c^2$, (b) $10.1 \, m_e c^2$, (c) $10.3 \, m_e c^2$ and (d) $10.1 \, m_e c^2$.}

\label{fig:component_contri}
\end{figure*}

\begin{figure}[ht!]\includegraphics[width=0.5\textwidth]{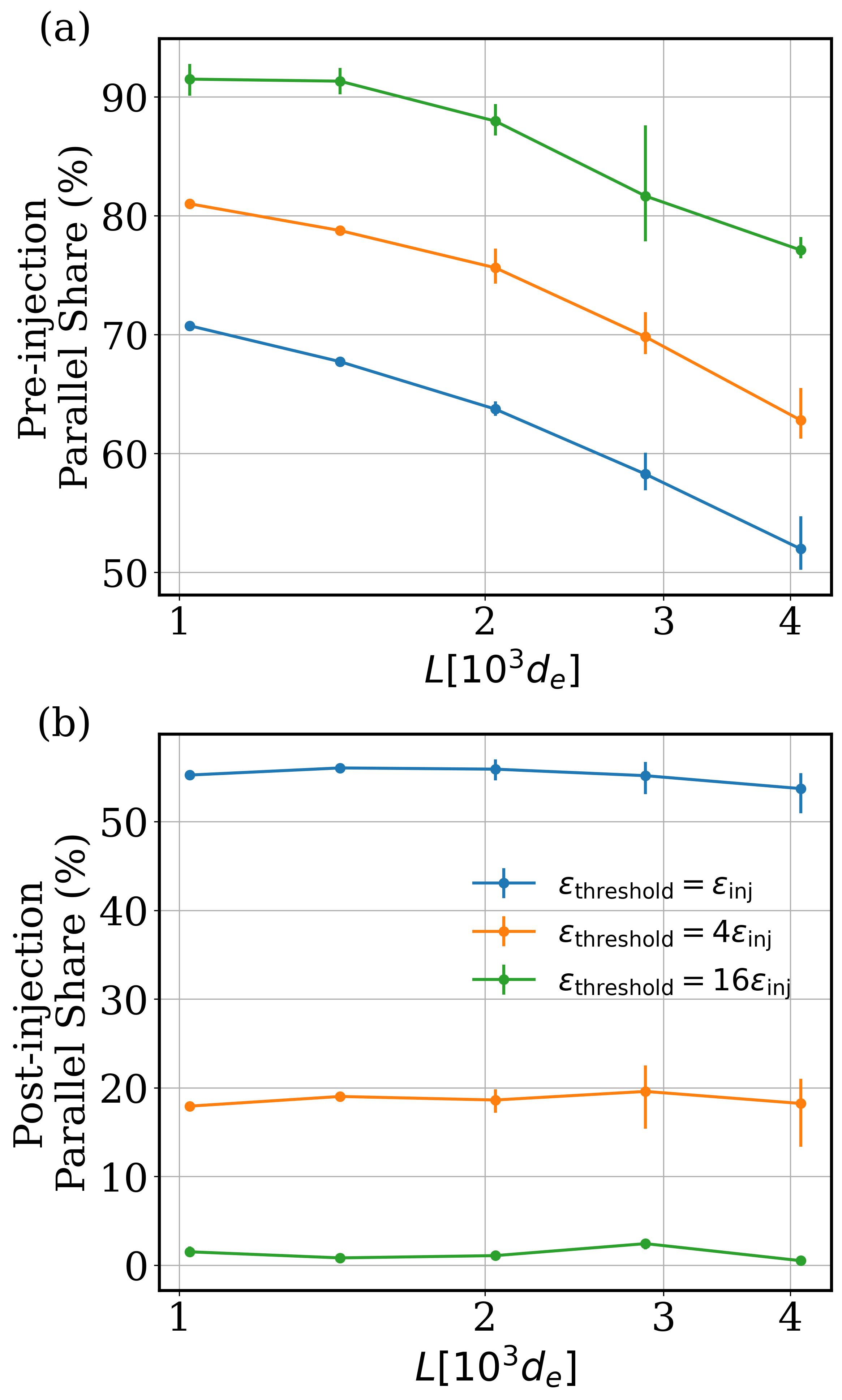}
\caption{Variation of (a) pre-injection and (b) post-injection share of the work done by the parallel electric field with domain size before and after injection for different $\varepsilon_{\rm threshold}$}. The plotted values are the weighted average of the simulations with seeds 1 and 2.
\label{fig:para_share}
\end{figure}

Figure~\ref{fig:spectra}(a) shows the time evolution of particle energy spectra for the simulation with domain size $L/d_e = 1440$. As the simulation starts, the turbulent magnetic fluctuations (Figure~\ref{fig:abs_j}) lead to strong particle acceleration and the development of a clear nonthermal power-law spectrum within $1$-$2$ light crossing times. The spectral index $p\sim 2.8$ and does not appreciably change in the late stage of the simulation. Figure~\ref{fig:spectra}(b) shows the nonthermal spectra obtained at final times for simulations with $L/d_e \in \{512, 1440, 4096\}$ (normalized to the total number of particles in each simulation). By performing the aforementioned fitting procedure on these spectra, we find that the injection energy~$\varepsilon_{\rm inj}$ is insensitive to the domain size~$L$, whereas the cutoff energy~$\varepsilon_c$ steadily increases with~$L$. The power-law index~$p$ steepens slightly with increasing domain size (see discussions below).

The spectral properties ($\varepsilon_{\rm inj}$, $\varepsilon_c$, $p$) are plotted against domain size~$L$ for all of our simulations in Figure~\ref{fig:spectrum_params}. We find that the simulation with~$L/d_e = 512$ was too small to yield precise measurements of these quantities (yielding a relatively large uncertainty), and therefore is not included. By inspecting Figure~\ref{fig:spectra}(b), we find the injection energy~$\varepsilon_{\rm inj}$ to be insensitive to domain size, the power-law index~$p$ to be slightly larger for larger domain sizes, and the cutoff energy~$\varepsilon_c$ to be larger for larger domain sizes, in accordance with the trends in Figure~\ref{fig:spectrum_params}. 

Figure~\ref{fig:spectrum_params}(a) shows that~$p$ only weakly depends on~$L$ and reaches~$p \simeq 2.9$ for the largest $L/d_e = 4096$, similar to \citet{Zhdankin_2018}. This weak dependence could be due to the decay of turbulence, leading to weaker acceleration in the late stage. The injection energy $\varepsilon_{\rm inj}$ shown in Figure~\ref{fig:spectrum_params}(b) follows a similar trend, converging around~$\varepsilon_{\rm inj} \simeq 10.5 \, m_e c^2$ ($\simeq (\sigma_0/2) m_e c^2$) with an error~$\pm \, 0.5 \, m_e c^2$. In contrast, $\varepsilon_c$ increases linearly with~$L$ (Figure~\ref{fig:spectrum_params}(c)), suggesting that particles can be accelerated to higher energies in simulations with larger domain sizes. 
Hence the power-law extent~$R$ grows linearly with increasing domain size (Figure~\ref{fig:spectrum_params}(d)), owing to the invariance of~$\varepsilon_{\rm inj}$ and linear rise of~$\varepsilon_c$ with increasing~$L$.

To better understand particle acceleration mechanisms, we analyze the energy gains of individual tracer particles and break them down into the work done by parallel ($W_\parallel$) and perpendicular ($W_\perp$) electric fields. This is done by first using the tracked particle data to calculate the electric field parallel to the local magnetic field $\bm{E_\parallel} = (\bm{E}\cdot \bm{B} /{B}^2) \bm{B}$ and perpendicular to it $\bm{E_\perp} = \bm{E} - \bm{E_\parallel}$. Then we can then calculate the work done by each component, i.e.~$W_\parallel(t) \equiv q\int_0^t \textbf{v}(t') \cdot \textbf{E}_\parallel(t') \,dt'$ and $W_\perp(t) \equiv q\int_0^t \textbf{v}(t') \cdot \textbf{E}_\perp(t') \,dt'$. 

Four examples of such tracer particles are shown in Figure~\ref{fig:component_contri}, with horizontal dashed lines indicating the injection \textit{threshold}~$W_{\rm inj} \equiv \varepsilon_{\rm inj} - \varepsilon_0$ for each particle, which represents the energy gain necessary for the particle to cross the injection energy~$\varepsilon_{\rm inj}$. Since the initial energy~$\varepsilon_0$ of each particle is sub-relativistic (i.e., $\lesssim 1$), the injection thresholds~$W_{\rm inj}$ hover just below the injection energy; in particular, $W_{\rm inj} \simeq $ (a) $9.5\, m_e c^2$, (b) $10.1\, m_e c^2$, (c) $10.3\, m_e c^2$ and (d) $10.1\, m_e c^2$ whereas~$\varepsilon_{\rm inj} \simeq 10.5$ for the case~$L/d_e = 1440$.

In Figure~\ref{fig:component_contri}(a), we see that for a high energy particle, the energy gain during injection is dominated by~$W_\parallel$. Later, $W_\parallel$ flattens out, and~$W_\perp$ dominates the energy gain. The pattern is similar to examples shown in \citet{Comisso2019} and has been seen in reconnection simulations \citep{Guo2015,Kilian2020,French_2023}. Hence, the subsequent acceleration for this particle to high energies is a result of the perpendicular electric fields via a Fermi-like mechanism. Figure~\ref{fig:component_contri}(b) shows a different high energy particle for which~$W_\parallel$ flattens out at a much lower energy and~$W_\perp$ dominates both the injection and post-injection phases. We also find relatively rare cases with~$W_\parallel$ dominating the post-injection phase, shown in Figure~\ref{fig:component_contri}(c) and (d).

Since every particle experiences a different evolution, our analysis is performed statistically over an ensemble of tracer particles (about 10-20\% of all the tracers) whose final energy exceeds~$\varepsilon_{\rm inj}$. Further, we monitor particles that cross certain energy thresholds $\varepsilon_{\rm threshold}$ separately. We break the energization process of each monitored particle into two phases: the energy gain up to the injection energy $\varepsilon_{\rm inj}$ termed \textit{pre-injection}, and subsequent energy gain termed \textit{post-injection}. The ``pre-injection parallel share" is defined as the fraction of monitored particles which have~$W_\parallel(t_{\rm inj}) > W_\perp(t_{\rm inj})$ (where~$t_{\rm inj}$ is the time step whereupon~$\varepsilon = \varepsilon_{\rm inj}$ is reached). 
Similarly, the ``post-injection parallel share" is defined as the fraction of monitored particles whose post-injection parallel energization exceeds perpendicular energization (i.e., $W_\parallel(t_{\rm final}) - W_\parallel(t_{\rm inj}) > W_\perp(t_{\rm final}) - W_\perp(t_{\rm inj})$, where~$t_{\rm final}$ is the final time step of the simulation). Figure~\ref{fig:para_share} shows the parallel share for particles with final energy~$\varepsilon_{\rm final} \geq \varepsilon_{\rm threshold} \in \{\varepsilon_{\rm inj}, 4\,\varepsilon_{\rm inj}, 16\,\varepsilon_{\rm inj}\}$.

We ran all of our simulations twice using the random number generator seeds to be 1 and 2. The values shown in Figure~\ref{fig:para_share} are the average of these two simulations and the error bars end points are the actual values of the two simulations.

For~$\varepsilon_{\rm threshold} = \varepsilon_{\rm inj}$ (blue line in Figure~\ref{fig:para_share}(a)), the pre-injection parallel share decreases with increasing domain size and drops to~$\sim 50\%$ for the largest domain, implying that~$W_\parallel$ and~$W_\perp$ play a comparable role in the initial particle energization. However, this curve has not yet saturated with increasing domain size, suggesting that~$W_\perp$ could dominate the injection stage for larger systems. As~$\varepsilon_{\rm threshold}$ increases, the pre-injection parallel share also increases. For very high energy particles ($\varepsilon_{\rm threshold} = 16\,\varepsilon_{\rm inj}$), the energy gain for most ($> 90\%$) particles is dominated by~$W_\parallel$ for small~$L$. For larger~$L$, the parallel share declines to~$\simeq 75\%$. This decreasing trend again indicates that the pre-injection parallel share fraction for high-energy particles could be even smaller for larger systems.

The post-injection shares are converged with system size~$L$ for each~$\varepsilon_{\rm threshold}$. For~$\varepsilon_{\rm threshold} = \varepsilon_{\rm inj}$, the parallel share is~$\sim 50\%$, indicating that~$W_\parallel$ and~$W_\perp$ contribute comparably to particle energization in the post-injection phase. As~$\varepsilon_{\rm threshold}$ increases, the post-injection parallel share decreases: When~$\varepsilon_{\rm threshold} = 4 \, \varepsilon_{\rm inj}$, $W_\parallel$ contributes~20\%, and for~$\varepsilon_{\rm threshold} = 16\,\varepsilon_{\rm inj}$, the~$W_\parallel$ contribution is negligible. This indicates that for very high energy particles, $W_\perp$ dominates the post-injection energy gain for almost all particles.

\section{Discussion and Conclusions} \label{sec:discussions_conclusion}

In this paper, we have presented results from 2D PIC simulations with~$\sigma_0 = 20$ and~$L/d_e$ varying from~$512$ to~$4096$ to investigate the mechanisms of nonthermal particle acceleration in turbulent plasma.

We find that for~$\varepsilon_{\rm threshold} = 16 \,\varepsilon_{\rm inj}$, the smaller domain sizes pre-injection parallel shares are higher than~$90\%$, indicating that $W_\parallel$ dominates the pre-injection phase for most particles. This is in alignment with the results of \cite{Comisso2019}, where they claim that initial particle acceleration is caused by $W_\parallel$. In the post-injection case for the same $\varepsilon_{\rm threshold}$, we find that the parallel share is close to~$0\%$, which indicates that almost all high energy particles get most of their energy from~$W_\perp$. This finding also aligns with \cite{Comisso2019}, which shows $W_\perp$ dominates late-stage energization. However, it must be noted that the particles analyzed by \cite{Comisso2019} are all very high energy with $\varepsilon_{\rm threshold} = 18 \sigma_0$. Even for high energy particles, we find that the pre-injection parallel share starts to decrease and drops to~$75\%$, indicating $W_\parallel$ only dominates the initial energization of three-quarters of the tracer particles. Given the decreasing trend continues at the largest box size (green line in Figure~\ref{fig:para_share}(a)), it is likely that the contribution by $W_\parallel$ in the pre-injection phase might be even smaller for astrophysical scale systems. Furthermore, when we look at the full picture by analyzing all injected tracer particles ($\varepsilon_{\rm threshold} = \varepsilon_{\rm inj}$), we recognize that $W_\perp$ plays a greater role in particle energization during the pre-injection phase, and $W_\parallel$ also a plays a more significant role in post-injection particle energization, especially particles with energy close to the lower bound of the power-law distribution.

We find strong agreement with \citet{Zhdankin_2018} in how the power-law index~$p$ depends on domain size~$L$ (c.f., Figure~\ref{fig:spectrum_params}). In particular, we find the power-law index to steadily steepen with increasing domain size, with~$p \simeq 2.9$ when~$L/d_e = 4096$. However it is still unclear at which domain size~$L/d_e$ and at what value~$p$ will converge. Simulations with continuous driving may help resolve this issue.

Our simulations use a constant magnetization~$\sigma_0=20$ and turbulence amplitude~${\delta B}_{\rm rms 0}/B_0=1$ in an electron-positron plasma. If the mechanisms that underlie injection in relativistic turbulence are the same as those for relativistic magnetic reconnection \citep{French_2023,Vega2024}, then the share of work done by~$E_\parallel$ ($E_\perp$) could increase with magnetization (c.f., Fig. 29 of \cite{Zhdankin2020}), but decrease with the turbulence amplitude. 
While electrons and positrons undergo identical injection processes, protons may undergo significantly different processes and requires a future study. Recent studies show that proton injection and acceleration in turbulence and magnetic reconnection are dominated by perpendicular electric field \citep{Comisso2022,Zhang2024apj}. Further studies are needed to resolve these important issues.

\begin{acknowledgments}
We acknowledge support through NSF Award 2308091, Los Alamos National Laboratory LDRD program, and DOE Office of Science. O.F. acknowledges support by the National Science Foundation Graduate Research Fellowship under Grant No. DGE 2040434. 
\end{acknowledgments}

\bibliography{biblio}{}
\bibliographystyle{aasjournal}

\end{document}